\documentclass[twocolumn,showpacs,preprintnumbers,amsmath,amssymb]{revtex4}
\begin{document}

\preprint{hep-th/0305121}

\title{Events in a Non-Commutative Space-Time.}
\author{M. Toller}
\email{toller@iol.it}
\homepage{http://www.science.unitn.it/~toller/}
\affiliation{via Malfatti n.\ 8, I-38100 Trento, Italy}
\date{\today}

\begin{abstract}
We treat the events determined by a quantum physical state in a non-commutative space-time, generalizing the analogous treatment in the usual Minkowski space-time based on positive-operator-valued measures (POVMs). We consider in detail the model proposed by Snyder in 1947 and calculate the POVMs defined on the real line that describe the measurement of a single coordinate. The approximate joint measurement of all the four space-time coordinates is described in terms of a generalized Wigner function (GWF). We derive lower bounds for the dispersion of the coordinate observables and discuss the covariance of the model under the Poincar\'e group. The unusual transformation law of the coordinates under space-time translations is interpreted as a failure of the absolute character of the concept of space-time coincidence. The model shows that a minimal length is compatible with Lorentz covariance.
\end{abstract}
\maketitle
  
\section{Introduction.}

It was recognized a long time ago \cite{Heisenberg} that there is no reason to believe that the usual space-time concepts maintain their validity at arbitrarily small scales of length and time. It has also been suggested that the interplay of quantum theory and general relativity does not permit the measurement of distances smaller than the Planck length and time intervals smaller than the Planck time \cite{Mead,Mead1,Ferretti,JR,AC2,Garay,Gibbs}. However, in the absence of a complete theory of quantum gravity, there is no agreement about the exact form of the limits to the validity of the classical space-time description.

A natural way to describe a space-time indeterminacy is to consider the space-time coordinates as elements of a non-commutative algebra \cite{Snyder,Yang,Golfand,Kadi,Golfand2,Cicogna,DFR,DFR1,KL}. This idea is suggested by the usual quantization procedure, which replaces the commutative algebra of the functions defined on the phase space by a non-commutative algebra of operators in Hilbert space. It has also been shown that a non-commutative space-time can be derived from a quantum deformation which replaces the enveloping algebra of the Poincar\'e Lie algebra by a non-cocommutative Hopf algebra (quantum group) \cite{MR,LRZ,KGN,KG}.

There are two different approaches to the quantization of space-time.
\begin{itemize}
\item One can build a mathematical structure, for instance a non-commutative algebra, which replaces the space-time manifold and the space-time coordinates. It is connected in some way with the algebra of the quantum observables, but it is not contained in it, in the same way as the classical space-time coordinates act as parameters, and not as observables, in a field theory.
\item One can study the quantum observables $X^{\alpha}$ interpreted as the space-time coordinates of an event defined by a physical object.
\end{itemize}
Both points of view are important and deserve attention, but one should carefully avoid any confusion between them. Note that in a theory based on the first point of view, one should still be able to define the coordinates of an event according to the second point of view. In many treatments one justifies the properties assumed for the coordinates interpèreted in the first way by means of physical arguments which concern the coordinates interpreted in the second way.

In the present paper we adopt the second approach, namely our aim is to locate physical events in space-time. By an event we mean a physical phenomenon which indicates, with some approximation, a time and three space coordinates. A typical example is the collision of two particles. In the center-of-mass system (disregarding for simplicity the quantum effects), one can identify the space coordinates of the event with the coordinates of the center-of mass and the time coordinate with the time at which the distance between the two particles takes its minimum value. In this way, the event is defined even in the absence of a close collision.

The definition of an event requires a well-defined physical system in a state described by a vector $\psi$ belonging to a Hilbert space $\mathcal{H}$. It admits a rigorous mathematical treatment and a clear physical interpretation.

A specific form for the space-time coordinate operators $X^{\alpha}$ in a theory symmetric with respect to the conformal group has been given in refs.\ \cite{JR1,JR2}. The mathematical nature of the coordinate operators concerning an event in a commutative Minkowski space-time, within a quantum theory symmetric with respect to an undeformed Poincar\'e group, has been discussed in refs.\ \cite{Toller,Toller1}. A problem arises because the coordinate operators $X^{\alpha}$, as a consequence of the support properties of the energy-momentum, cannot be self-adjoint \cite{Pauli,Wightman}. We shall find the same problem in the non-commutative case. 

A completely satisfactory solution of this problem is obtained by replacing the spectral measure corresponding to a self-adjoint operator by a positive-operator-valued measure (POVM) \cite{Davies,Holevo,Werner,BGL1}. The same idea permits a correct treatment of the time observable and of the ``time of arrival'' relevant for  time of flight measurements (see for instance \cite{BGL,Giannitrapani,MLP}). 

It has been shown in ref.\ \cite{Toller1} that the non-self-adjointness of the coordinate operators gives rise to uncertainty relations stronger than the ones which follow in the usual way from the commutation relations \cite{Messiah}. These effects have to be taken into account in a discussion of the properties of a non-commutative space-time.

It is important to remark that the ideas outlined above and developed in the following sections have a provisional character, because the coordinates are measured with respect to a classical frame of reference, which is an idealized concept. One should consider quantum reference frames \cite{AK,Rovelli,Toller2,Mazzucchi}, described by physical quantum objects. Then, the velocity and the angles which determine the orientation of these objects come into play, together with the space-time coordinates of the origin, and have to be quantized. 

This program, which lies outside the scope of the present paper, has two steps: first one has to quantize the parameters which determine the relation between a quantum frame and a classical frame, then one has to consider the relation between two quantum frames. One may say that the problem is to quantize the Poincar\'e group, but it seems \cite{Toller2} that the solution is not a quantum group in the usual sense, namely a Hopf algebra.

In the next section, in order to present the necessary mathematical tools, we summarize the treatment of quantum events in commutative Minkowski space-time. In section III we introduce the Snyder model of non-commutative space-time and we define the coordinate observables in terms of POVMs on the real line and of a generalized Wigner function in the classical space-time. In section IV and V we develop the formalism obtaining more explicit formulas. 

In section VI we discuss the symmetry under space-time translations, which, in the model we are considering, presents rather unusual features. In section VII we calculate some lower bounds to the variance of the coordinate observables, justifying the initial motivations of the model. In section VIII we show how the approximate joint measurement of the four coordinates can be treated by generalizing the Wigner function formalism used to treat the approximate joint measurement of the non-commuting coordinates of the phase space \cite{Wigner2,LT,Busch}. In section IX we summarize the main results.

\section{Events in a commutative Minkowski space-time.}

In this section we summarize the treatment of quantum events in the ordinary Minkowski space-time, omitting unnecessary details and emphasizing the steps to be modified when a non-commutative space-time is considered.  Instead of the treatment of ref.\ \cite{Toller}, we follow the more elegant approach given in ref.\ \cite{Mazzucchi}, based on the ideas of ref.\ \cite{Werner}.

The event is determined by a physical object described by a vector $\psi$ belonging to the Hilbert space $\mathcal{H}$ in which a unitary representation
\begin{equation} 
U(a, \Lambda) = T(a)V(\Lambda), \qquad a \in \mathcal{T}, \qquad \Lambda \in \mathcal{L}
\end{equation} 
of the proper orthochronous Poincar\'e group $\mathcal{P}$ operates \cite{Wigner}. $\mathcal{T}$ is the space-time translation group, $\mathcal{L}$ is the proper orthochronous Lorentz group, $a$ is a four-vector and  $\Lambda$ is a Lorentz $4 \times 4$ matrix. We consider only states with integral angular momentum, but all our arguments can easily be extended to states with half-integral angular momentum by introducing the universal covering $SL(2, \mathbf{C})$ of $\mathcal{L}$.

The translation unitary operators have the spectral representation
\begin{equation} \label{Trans}
T(a) = \int_{\mathcal{Q}} \exp(i p \cdot a) \, d\mu(p),
\end{equation}
where $\mathcal{Q} = \mathcal{T}^*$ is the four-momentum space,  $p \in \mathcal{Q}$  and $p \cdot a = p^0 a^0 - \mathbf{p} \cdot \mathbf{a}$ is the relativistic scalar product of two four-vectors. $\mu$ is a spectral measure, which assigns to a Borel set $J \subset \mathcal{Q}$ a projection operator $\mu(J)$ in a numerably additive way.  One can show that
\begin{equation} \label{Impr}
V(\Lambda)\mu(J)V^{-1}(\Lambda) = \mu(\Lambda J),
\end{equation}
where $\Lambda J$ is the set $J$ transformed by the Lorentz matrix $\Lambda$. According to Mackey \cite{Mackey,Mackey1,Mackey2}, the unitary representation $V(\Lambda)$ and the spectral measure $\mu(J)$ satisfying the relation (\ref{Impr}) form an \textit{imprimitivity system}, which we indicate by $(V, \mu)$.  

If the spectral measure $\mu$ is concentrated in an orbit $\mathcal{O} \subset \mathcal{Q}$, we say that the imprimitivity system is \textit{transitive}. In this case, according to Mackey's imprimitivity theorem, the representation $V$ can be described explicitly as an \textit{induced representation}, which is exactly the one obtained in Wigner's fundamental paper \cite{Wigner}. In this way one obtains all the irreducible unitary representation of the Poincar\'e group and some of them describe the ``elementary'' particles.

We assume asymptotic completeness \cite{SW}, namely that all the physical states can be described in terms of \textit{in} or \textit{out} particle states. We consider the direct sum decomposition  $\mathcal{H} = \mathcal{H}_S \oplus \mathcal{H}_C$, where $\mathcal{H}_S$ contains states with a discrete mass spectrum, namely the vacuum and the one-particle states, while $\mathcal{H}_C$ contains the states with a continuous mass spectrum, namely the many-particle states.  It is physically evident, and it follows from the formalism \cite{Toller}, that the vacuum and the one-particle states cannot define an event. They are too simple to be treated as a ``clock'' which determines the time coordinate. Thus, in our treatment of the events, we consider only states belonging to $\mathcal{H}_C$ and we write $\mathcal{H}$  instead of $\mathcal{H}_C$.

In order to describe many-particle states, we have to consider non-transitive imprimitivity systems. The support $\mathcal{V}$ of the measure $\mu$ is composed of many orbits and is contained in the closed future cone.  It is given by
\begin{equation} \label{Future}
s = p \cdot p \geq s_0 = (2 m_0)^2, \qquad p^0 \geq 0,
\end{equation}
where $m_0$ is the smallest of the particle masses (possibly vanishing).

The Hilbert space $\mathcal{H}$ is decomposed into a direct integral of spaces in which irreducible unitary representations (IURs) of  $\mathcal{P}$ operate. Of course, only positive-energy representations appear in this decomposition. Since we are not considering one-particle states, we can disregard zero-mass representations and we consider only positive-mass IURs, which are labeled by the mass $s^{1/2}$ and the center-of-mass angular momentum $j$.  

A vector $\psi \in \mathcal{H}$ is described by a wave function of the kind $\psi_{\sigma j m}(p)$, where $m = -j, -j + 1,\cdots, j$ describes the third component of the center-of-mass angular momentum and the index $\sigma$ stands for all the other quantum numbers. For instance, in a two-particle state $\sigma$ describes the center-of-mass helicities \cite{JW}. It is not necessary to specify the mass $s^{1/2}$, since it is a function of $p$. The range of the indices $j, \sigma$ may depend on $s$. For fixed $\sigma$, $s$ and $j$, the group  $\mathcal{P}$ acts according to the induced representation described by Wigner \cite{Wigner}. 

The norm and the action of the translation group are given by 
\begin{equation}  \label{Norm}
\|\psi\|^2 = 
\int_{\mathcal{V}} \sum_{\sigma j m} |\psi_{\sigma j m}(p)|^2  \, d^4 p,
\end{equation}
\begin{equation} \label{Trans1}
[T(a)\psi]_{\sigma j m}(p) = \exp(ip \cdot a) \psi_{\sigma j m}(p). 
\end{equation}
We choose in each orbit a representative element $\hat p(s)$ and for each four-momentum $p \in \mathcal{V}$ an element $\Lambda_p \in \mathcal{L}$ with the property 
\begin{equation} \label{Boosts}
p = \Lambda_p \hat p(s), \qquad  \hat p(s) = (s^{1/2}, 0, 0, 0)^T,
\qquad s = p \cdot p.
\end{equation}
The operator $V(\Lambda)$ is defined by
\begin{equation} \label{Ind}
[V(\Lambda)\psi]_{\sigma j m}(p) = R^j_{mm'}(\Theta) \psi_{\sigma j m'}(p'), 
\end{equation}
where
\begin{equation} 
p' = \Lambda^{-1} p, \qquad  \Theta = \Lambda_p^{-1} \Lambda \Lambda_{p'} \in SO(3)
\end{equation}  
and $R^j_{mm'}(\Theta)$ is the $(2j +1)$-dimensional IUR of $SO(3)$.

According to a naive application of the rules of quantum mechanics, the commuting operators $X^{\alpha}$, which represent the coordinates of an event, should have a joint spectral representation
\begin{equation} \label{Coo}
X^{\alpha} = \int_{\mathcal{M}} x^{\alpha} \, d\tau(x),
\end{equation}
where $\tau$ is a spectral measure on the Minkowski space-time $\mathcal{M}$, and $x^{\alpha}$ are the (numerical) coordinates of this space. If $\psi \in \mathcal{H}$ with $\|\psi\| = 1$ defines a state of the system, the quantity $(\psi, \tau(I) \psi)$ is the probability that the results of a joint measurement of the four coordinates define a point belonging to the Borel set $I \subset \mathcal{M}$.

A physical requirement is the Poincar\'e covariance, given by the condition
\begin{equation} \label{Cov}
U(a, \Lambda)\tau(I)U^{-1}(a, \Lambda) = \tau(\Lambda I + a).
\end{equation}
This equation means that the representation $U(a, \Lambda)$ of $\mathcal{P}$ and the spectral measure $\tau$ on $\mathcal{M}$ form a transitive imprimitivity system, which we indicate by $(U, \tau)$.

However, it is known that the equations given above lead to a contradiction with the properties of the energy-momentum spectrum \cite{Wightman,Toller}. In fact, the unitary operators $\exp(-ib \cdot X)$ describe translations in the space $\mathcal{Q}$, which lead to states with unphysical energy-momentum. This problem is avoided by assuming that the operators $\tau(I)$ are not projection operators, but just positive bounded operators. In other words, $\tau$ is a positive-operator valued measure (POVM). The physical meaning of $(\psi, \tau(I) \psi)$ is unchanged and the description of quantum observables in terms of POVMs is perfectly compatible with the standard interpretation of quantum mechanics \cite{Davies,Holevo,Werner,BGL1}.  

It was shown in ref.\ \cite{Giannitrapani1} that, the operator $\tau(I)$ cannot represent a quasi-local observable \cite{Haag}. In particular, this observable cannot be measured by means of operations performed exclusively in the space-time region $I$.

The coordinate operators (\ref{Coo}) are Hermitian, but not self-adjoint. The covariance equation (\ref{Cov}) is still valid, but, instead of an imprimitivity system, we have a \textit{covariance system}, still indicated by $(U, \tau)$, and the imprimitivity theorem cannot be applied.

Great help comes from a theorem \cite{Werner,Scutaru,Cattaneo,CH} which asserts that a covariance system can always be obtained from an imprimitivity system, which, under some conditions, is unique up to isomorphisms. In our case, we can find an imprimitivity system, indicated by  $(\tilde U, \tilde\tau)$, formed by the representation $(a, \Lambda) \to \tilde U(a, \Lambda) = \tilde T(a) \tilde V(\Lambda)$ of $\mathcal{P}$ and a spectral measure $\tilde \tau$, on the space-time $\mathcal{M}$, both acting in the auxiliary Hilbert space $\tilde{\mathcal{H}}$ and satisfying the covariance condition
\begin{equation} \label{Cov2}
\tilde U(a, \Lambda) \tilde\tau(I) \tilde U^{-1}(a, \Lambda) = \tilde\tau(\Lambda I + a).
\end{equation}

The connection with the covariance system is given by
\begin{equation} \label{Intertwin}
\tau(I) = A^{\dagger} \tilde \tau(I) A, \qquad
A U(a, \Lambda) = \tilde U(a, \Lambda) A, 
\end{equation}
where $A$ is a bounded linear mapping from $\mathcal{H}$ to $\tilde{\mathcal{H}}$. The last equation means that it is an \textit{intertwining operator} between the representations $U$ and $\tilde U$.  If we assume that the event necessarily takes place somewhere in space-time, we have $\tau(\mathcal{M}) = 1$, and it follows
\begin{equation} \label{Isom}
A^{\dagger} A = 1,
\end{equation}
namely $\mathcal{H}$ is mapped isometrically onto a subspace of $\tilde{\mathcal{H}}$. The transitive imprimitivity system $(\tilde U, \tilde\tau)$ can be treated by means of the imprimitivity theorem and one finds the explicit form of the representation $\tilde U$ and of the auxiliary space $\tilde{\mathcal{H}}$. The complete treatment of the intertwining operator $A$ and of the POVM $\tau$ can be found in refs. \cite{Toller,Toller1,Mazzucchi}.

If we introduce the self-adjoint operators
\begin{equation} \label{Xtilde}
\tilde X^{\alpha} = \int_{\mathcal{M}} x^{\alpha} \, d \tilde \tau(x),
\end{equation}
the Hermitian coordinate operators are given by
\begin{equation} \label{XXTilde}
X^{\alpha} = A^{\dagger} \tilde X^{\alpha} A.
\end{equation}

The commuting operators $\tilde T(a)$ have the spectral representation
\begin{equation} \label{Trans2}
\tilde T(a) = \int_{\mathcal{Q}} \exp(i p \cdot a) \, d \tilde \mu(p)
\end{equation}
where the spectral measure $\tilde \mu$ is Lorentz covariant, namely, in analogy with eq.\ (\ref{Impr}), we have
\begin{equation} \label{Cov3}
\tilde V(\Lambda) \tilde\mu(J) \tilde V^{-1}(\Lambda) = \tilde\mu(\Lambda J).
\end{equation}
By comparing eq.\ (\ref{Trans2}) with eqs.\ (\ref{Trans}) and (\ref{Intertwin}) we obtain
\begin{equation} \label{AMu}
A \mu(J) = \tilde \mu(J) A, \qquad J \subset \mathcal{Q}.
\end{equation}
We can also consider the unphysical energy-momentum operators
\begin{equation}
\tilde P^{\alpha} = \int_{\mathcal{Q}} p^{\alpha} \, d \tilde \mu(p)
\end{equation}
defined in the auxiliary space $\tilde{\mathcal{H}}$.

We introduce the unitary operators
\begin{equation} 
\tilde W(b) = \int_{\mathcal{M}} \exp(- i b \cdot x) \, d \tilde \tau(x)
\end{equation}
and from eq.\ (\ref{Cov2}) it follows that
\begin{equation} \label{Weyl}
\tilde T(a) \tilde W(b) \tilde T(-a) = \exp(i a \cdot b) \tilde W(b).
\end{equation}
This equation shows that the operators $\tilde T(a) = \exp(i a \cdot \tilde P)$ and  $\tilde W(b) = \exp(-i b \cdot \tilde X)$ form a unitary representation of the four-dimensional Weyl-Heisenberg group \cite{Weyl}, which is a precise formulation of the canonical commutation relations 
\begin{equation} 
[\tilde P^{\alpha}, \tilde X^{\beta}] = i g^{\alpha \beta}.
\end{equation}
These operators, however, do not operate in the physical Hilbert space $\mathcal{H}$, but in the auxiliary space $\tilde{\mathcal{H}}$.

It follows from eq.\ (\ref{Weyl}) that
\begin{equation} \label{Cov4}
\tilde W(b)\tilde \mu(J) \tilde W(-b) = \tilde \mu(J + b).
\end{equation}
This covariance equation shows that the unitary operators $\tilde W(b)$ describe translations in the energy-momentum space. It follows that the support of $\tilde \mu$ is the whole energy-momentum space $\mathcal{Q}$, but there is no problem, because the spectrum of $\tilde \mu$ is not the physical energy-momentum spectrum. The generators of these translations are the self-adjoint operators $\tilde X^{\alpha}$.

In conclusion, we have studied a large group acting unitarily on the auxiliary space $\tilde{\mathcal{H}}$. It is the semi-direct product of the Lorentz group $\mathcal{L}$ and the four-dimensional Weyl-Heisenberg group, which, in turn, contains the space-time translation group $\mathcal{T}$ and the translation group $\mathcal{T}'$ of the energy-momentum space. Of course, $\mathcal{T}'$ cannot act on the physical Hilbert space $\mathcal{H}$. We indicate by $\mathcal{P}'$ the subgroup generated by $\mathcal{L}$ and $\mathcal{T}'$. It is isomorphic to the usual Poincar\'e group $\mathcal{P}$, but its meaning is different.

The operators $S(b, \Lambda) = \tilde W(b)\tilde V(\Lambda)$, defined in $\tilde{\mathcal{H}}$ form a unitary representation of the group $\mathcal{P}'$. From eqs.\ (\ref{Cov3}) and (\ref{Cov4}), we see that this representation, together with the spectral measure $\tilde \mu$ defined on $\mathcal{Q}$, forms an imprimitivity system $(S, \tilde \mu)$, which is the starting point of the next section.  
 
\section{The Snyder model.}

Now we are ready to discuss the model of non-commutative space-time proposed a long time ago by Snyder \cite{Snyder,Yang,Golfand,Kadi,Golfand2,Cicogna}. Actually, as we shall see, there are several options and it is more correct to speak of a class of Snyder's models. Recently it has been shown that some of them are related to models obtained from quantum Poincar\'e groups \cite{KGN,KG}. Note that, while the formalism summarized in the preceding section is derived in a univocal way from sound physical principles, the modifications considered in the following are just a provisional attempt, based on Snyder's ideas and on the analogy with the commutative case.

The idea, reformulated in our language, is to replace the imprimitivity system $(S, \tilde\mu)$ introduced at the end of the preceding Section, by another imprimitivity system, denoted in the same way, where the group $\mathcal{P}'$ is replaced by another group $\mathcal{G}$, containing the Lorentz group $\mathcal{L}$, and $S$ is a unitary representation of $\mathcal{G}$ acting on the auxiliary Hilbert space $\tilde{\mathcal{H}}$. The action of the Poincar\'e group $\mathcal{P}$ on the space $\mathcal{H}$ of the physical states is still described by the unitary representation $U$ and by the imprimitivity system $(V, \mu)$ introduced in the preceding section.  

The manifold $\mathcal{Q}$, on which the spectral measure $\tilde\mu$ is defined is a homogeneous space of $\mathcal{G}$ and it has to be modified with respect to the one introduced in section II. It must contain a Lorentz invariant set $\mathcal{V}$, identified with the support of the physical spectral measure $\mu$, in such a way that eq.\ (\ref{AMu}) is still meaningful. We call $\mathcal{Q}$ the \textit{extended energy-momentum space}.

As in the preceding Section, we indicate by $\tilde X^{\alpha}$ the self-adjoint generators of four suitably chosen one-parameter subgroups of $\mathcal{G}$. Now, however, these operators do not necessarily commute. The operators which represent the coordinate observables are defined by eq.\ (\ref{XXTilde}). As in the commutative case, and for the same reasons, it is not possible to use directly the operators $\tilde X^{\alpha}$, which do not operate in the physical Hilbert space. 

The natural choices for the group $\mathcal{G}$ are the connected components of the identity of the de Sitter group $SO(1, 4)$ or the anti-de Sitter group  $SO(2, 3)$, or their universal coverings. We shall treat the first choice in detail, but the other cases can be treated in a similar way. By considering the universal covering, one can also treat events defined by systems with half-integral angular momentum.

We consider $\mathcal{G}$ as a group of real matrices operating on a five-dimensional vector space with coordinates $\xi^{\mu}$ and a diagonal metric tensor $g_{\mu \nu}$ defined by $g_{00} = 1$, $g_{11} = g_{22} = g_{33} = g_{55} = - 1$. Here and in the following, the indices  $\mu, \nu, \rho, \sigma$ take the values  $0, 1, 2, 3, 5$, while the indices $\alpha, \beta, \gamma$ take the values  $0, 1, 2, 3$. The matrices $\Gamma \in \mathcal{G}$ satisfy the condition
\begin{equation} 
\Gamma^T g  \Gamma = g.
\end{equation}
The matrices that do not affect the coordinate $\xi^5$ form the Lorentz subgroup $\mathcal{L}$. In the following, we indicate by the same symbol a $4 \times 4$ Lorentz matrix and the corresponding $5 \times 5$ matrix belonging to $\mathcal{G}$.

The infinitesimal transformations are represented by the matrices $\Xi_{\rho \sigma} = - \Xi_{\sigma \rho}$ defined by
\begin{equation} 
\Gamma^{\mu}{}_{\nu} = \delta^{\mu}_{\nu} + \epsilon^{\mu}{}_{\nu} + O(\epsilon^2) =
\delta^{\mu}_{\nu} + \frac 1 2  \epsilon^{\rho \sigma} \Xi_{\rho \sigma}{}^{\mu}{}_{\nu}
+ O(\epsilon^2),
\end{equation}
\begin{equation} 
\Xi_{\rho \sigma}{}^{\mu}{}_{\nu} =
\delta^{\mu}_{\rho} g_{\nu \sigma} - \delta^{\mu}_{\sigma} g_{\nu \rho}.
\end{equation}
They satisfy the commutation relations
\begin{equation} 
[\Xi_{\mu \nu}, \Xi_{\rho \sigma}] = 
g_{\nu \rho} \Xi_{\mu \sigma} 
- g_{\mu \rho} \Xi_{\nu \sigma}
- g_{\nu \sigma} \Xi_{\mu \rho} + g_{\mu \sigma} \Xi_{\nu \rho}
\end{equation}
of the Lie algebra $o(1, 4)$.

It follows that the self-adjoint generators $\tilde M_{\rho \sigma}$ of the unitary representation $S(\Gamma)$ defined by
\begin{equation} \label{Inf}
S\left(\exp (2^{-1} \epsilon^{\rho \sigma} \Xi_{\rho \sigma})\right) =
\exp\left(- 2^{-1} i \epsilon^{\rho \sigma} \tilde M_{\rho \sigma}\right)
\end{equation}
satisfy the commutation relations
\begin{eqnarray}
&[\tilde M_{\mu \nu}, \tilde M_{\rho \sigma}]& 
\nonumber \\ \label{Commut}
&= i (g_{\nu \rho} \tilde M_{\mu \sigma} - g_{\mu \rho} \tilde M_{\nu \sigma}
- g_{\nu \sigma} \tilde M_{\mu \rho} + g_{\mu \sigma} \tilde M_{\nu \rho}). \qquad &
\end{eqnarray}

We put 
\begin{equation} \label{Coord}
\tilde X^{\alpha} =l \tilde M^{\alpha 5},
\end{equation} 
where $l$ is a fundamental length, and we obtain
\begin{equation} \label{CommX}
[\tilde X_{\alpha},\tilde  X_{\beta}] =  i l^2 \tilde M_{\alpha \beta},
\end{equation}
\begin{equation} 
[\tilde M_{\alpha \beta}, \tilde X_{\gamma}] = 
i (g_{\beta \gamma} \tilde X_{\alpha} - g_{\alpha \gamma} \tilde X_{\beta}).
\end{equation} 
The last formula shows that the operators $\tilde X^{\alpha}$ transform as the components of a four-vector under the action of the Lorentz group. If, in agreement with the preceding Section, we put
\begin{equation} 
\tilde V(\Lambda) = S(\Lambda), \qquad \Lambda \in \mathcal{L},  
\end{equation}
we obtain the following  transformation property of $\tilde X^{\alpha}$ under finite Lorentz transformations
\begin{equation} \label{FinLor}
\tilde V(\Lambda^{-1}) \tilde X^{\alpha} \tilde V(\Lambda) = \Lambda^{\alpha}{}_{\beta} \tilde X^{\beta}.
\end{equation}
Note that, since the operators $\tilde X^{\alpha}$ do not commute, one cannot write for them a joint spectral representation of the kind (\ref{Xtilde}).

If we abandon the Lorentz covariance, still maintaining the rotational covariance, we have more freedom and, for instance, as suggested in refs. \cite{KGN,KG}, we can replace the space components of eq.\  (\ref{Coord}) by the expressions
\begin{equation} \label{Coord2}
\tilde X^r = l (\tilde M^{r 5} + \tilde M^{r 0}), \qquad r = 1, 2, 3,
\end{equation} 
and we have the commutation relations
\begin{equation} \label{CommX2}
[\tilde X^r,\tilde  X^s] = 0, \qquad
[\tilde X^0,\tilde  X^r] = -i l \tilde X^r,
\end{equation}
which were derived in \cite{MR} from the quantum group formalism. In the following we shall adopt the definition (\ref{Coord}), but a model based on eq.\ (\ref{Coord2}) can be treated in a similar way. 

As we anticipated above, the physical coordinates of the event are described by the operators $X^{\alpha}$ defined in the physical Hilbert space $\mathcal{H}$ by eq.\ (\ref{XXTilde}), where the intertwining operator $A: \mathcal{H} \to \tilde{\mathcal{H}}$ has the properties (\ref{Isom}), (\ref{AMu}) and
\begin{equation} \label{LorCov}
A V(\Lambda) = \tilde V(\Lambda) A, \qquad \Lambda \in \mathcal{L}.
\end{equation}
The condition (\ref{AMu}) is related to the translational symmetry of the operator $A$. This delicate problem will be discussed in section V and VI.

In order to get a detailed physical interpretation, it is not sufficient to know the Hermitian operators $X^{\alpha}$, because they do not determine uniquely the corresponding POVMs defined on the real line $\mathbf{R}$. It is also interesting to consider more general observables of the kind 
\begin{equation} 
k \cdot X = k_{\alpha} X^{\alpha} = A^{\dagger} k \cdot \tilde X A.
\end{equation}
Since the operator $k \cdot \tilde X$ is the generator of a one-parameter subgroup of $\mathcal{G}$, it is self-adjoint and it has the spectral representation
\begin{equation} \label{SpectX}
k \cdot \tilde X = \int_{\mathbf{R}} \lambda \, d \tilde\tau_k(\lambda),	
\end{equation}
which defines the spectral measure $\tilde\tau_k$ implicitly.
The statistics of the results of a measurement of $k \cdot X$ is completely described by the POVM
\begin{equation} \label{POVM}
\tau_k(I) = A^{\dagger} \tilde\tau_k(I)  A, \qquad I \subset \mathbf{R}.
\end{equation}

We have, as usual, 
\begin{equation} 
\langle k \cdot x\rangle  = 
\left(\psi, A^{\dagger} \int_{\mathbf{R}} \lambda \, d \tilde\tau_k(\lambda) A \psi \right) = 
(\psi, k \cdot X \psi).
\end{equation}
However, for the square and higher powers of the coordinates, we obtain a more complicated expression, namely
\begin{eqnarray} 
&\langle (k \cdot x)^2 \rangle  = 
\left(\psi, A^{\dagger} \int_{\mathbf{R}} \lambda^2 \, d \tilde\tau_k(\lambda) A \psi \right)&
\nonumber \\ \label{XQuadr}
&= \left(\psi, A^{\dagger} (k \cdot  \tilde X)^2 A \psi \right)
\geq \left(\psi, (k \cdot X)^2 \psi \right).&
\end{eqnarray}
From these formulas we obtain the variances
\begin{equation} \label{Var}
\Delta x^{\alpha} = \left(\langle (x^{\alpha})^2 \rangle - \langle x^{\alpha}\rangle^2 \right)^{1/2}.
\end{equation}
The corresponding uncertainty relations are discussed in section VII. 

More generally, we may consider the average value
\begin{eqnarray} 
&\langle f(k \cdot x) \rangle = (\psi, \int f(\lambda) \, d\tau_k(\lambda)\,\psi)&
\nonumber \\
&= \int \tilde f(\theta) \langle \exp(-i \theta k \cdot x)\rangle \, d\theta,&
\end{eqnarray}
where
\begin{eqnarray} 
&\tilde f(\theta) = (2 \pi)^{-1}
\int f(\lambda) \exp(i \theta \lambda) \, d\lambda,& 
\nonumber \\
&f(\lambda) = \int \tilde f(\theta) \exp(-i \theta \lambda) \, d\theta,&
\end{eqnarray}
\begin{eqnarray} 
&\langle \exp(-i \theta k \cdot x)\rangle =
\left(\psi, A^{\dagger} \int \exp(-i \theta \lambda) \, d\tilde\tau_k(\lambda)\, A \psi \right)&
\nonumber \\
&= \left(\Psi, \exp(-i \theta  k \cdot  \tilde X) \Psi \right) =
\left(\Psi, S(\exp(-\theta l k^{\alpha} \Xi_{\alpha 5})) \Psi \right), &
\nonumber \\
&\Psi = A \psi. &
\end{eqnarray} 

If we introduce the real function
\begin{equation}  \label{Wig}
\rho(x) = (2 \pi)^{-4}\int \exp(i k\cdot x) 
\left(\Psi, S(\exp(-l k^{\alpha} \Xi_{\alpha 5})) \Psi \right) \, d^4 k,
\end{equation} 
we obtain
\begin{equation} 
\langle f(k \cdot x) \rangle = 
\int f(k \cdot x) \rho(x) \, d^4 x.
\end{equation} 
Note, however, that $\rho(x)$ cannot be interpreted as a probability density in the Minkowski space-time $\mathcal{M}$, because it may take negative values. It plays the same role as the Wigner function in the phase space \cite{Wigner2} and we call it the generalized Wigner function (GWF). It is the Fourier transform of a continuous bounded function and since, at the present stage, we do not know if this function decreases sufficiently fast at infinity, $\rho$ may be a generalized function in the sense of ref. \cite{GGV}, namely a distribution. 

The GWF $\rho(x)$ is completely determined by the average values $\langle f(k \cdot x) \rangle$, namely it can be measured with any required accuracy by performing many coordinate measurements on many states prepared in the same way. It describes the statistical properties of the observables $X^{\alpha}$ completely and provides a useful tool independently of the model considered. 

We can also write
\begin{equation} 
\rho(x) = 
(\psi, \tau(x) \psi),
\end{equation} 
where
\begin{eqnarray} 
&\tau(x) = (2 \pi)^{-4}\int \exp(i k \cdot x) 
A^{\dagger} S(\exp(-l k^{\alpha} \Xi_{\alpha 5})) A \, d^4 k&
\nonumber \\
&= (2 \pi)^{-4}\int \exp(i k \cdot x) 
A^{\dagger} \exp(-i k \cdot \tilde X) A \, d^4 k&
\end{eqnarray} 
is an Hermitian operator-valued distribution on the Minkowski space-time, that replaces the POVM $\tau$ in the non-commutative case. The Hermiticity follows from the unitarity of $S$. It is the Fourier transform of a bounded continuous operator-valued function. 

From eqs.\ (\ref{FinLor}) and (\ref{LorCov}) we obtain the Lorentz covariance property
\begin{equation}
V(\Lambda) \tau(x) V^{-1}(\Lambda) = \tau(\Lambda x).
\end{equation}

\section{The induced representation $S$.}

Now we have to enter into the details of the formalism proposed in the preceding section. The extended energy-momentum space $\mathcal{Q}$ is an orbit in $\mathbf{R}^5$ defined by the condition
\begin{equation} 
g_{\mu \nu} \xi^{\mu} \xi^{\nu} = (\xi^0)^2 - \|\boldsymbol{\xi}\|^2 - (\xi^5)^2  = - 1,
\end{equation}
where $\boldsymbol{\xi}$ is a vector with components $\xi^1, \xi^2, \xi^3$. We indicate by $p$ a generic element of $\mathcal{Q}$. The quantities $\xi^{\mu}$ form a redundant system of coordinates in the manifold $\mathcal{Q}$, which have no particular physical meaning. An invariant measure on this manifold is given by
\begin{eqnarray} 
&d \nu(p) = 2 l^{-4} \delta(\xi_{\mu} \xi^{\mu} + 1 ) d^5 \xi&
\nonumber \\
&= l^{-4} |\xi^5|^{-1} d\xi^0 d^3\boldsymbol{\xi} =  
l^{-4} |\xi^0|^{-1} d\xi^5 d^3\boldsymbol{\xi}. &
\end{eqnarray}

In order to write the induced representation corresponding to the transitive imprimitivity system $(S, \tilde\mu)$, we choose the element $\hat p \in \mathcal{Q}$ with coordinates $\boldsymbol{\xi} = 0$, $\xi^0 = 0$, $\xi^5 = 1$. The corresponding stability group is the Lorentz group $\mathcal{L}$, considered as a subgroup of $\mathcal{G}$. The inducing representation $\Lambda \to D(\Lambda)$ is a unitary representation of $\mathcal{L}$ and we indicate by $\hat{\mathcal{H}}$ the Hilbert space in which it operates. We choose the elements $\Gamma_p \in \mathcal{G}$ with the property
\begin{equation} \label{Boosts2}
p = \Gamma_p \hat p, \qquad p \in \mathcal{Q}.
\end{equation}

The Hilbert space $\tilde{\mathcal{H}}$ is composed of functions $\Psi(p)$ defined on $\mathcal{Q}$ with values in the Hilbert space $\hat{\mathcal{H}}$ and with norm given by
\begin{equation} 
\|\Psi\|^2 = \int_{\mathcal{Q}} \|\Psi(p)\|^2 \, d\nu(p).
\end{equation}
The induced representation $S(\Gamma)$ is defined by
\begin{equation} \label{Ind2}
[S(\Gamma)\Psi](p) = D(\Lambda) \Psi(p'), 
\end{equation}
where
\begin{equation}  \label{IndDef}
p' = \Gamma^{-1} p, \qquad \Lambda = \Gamma_p^{-1} \Gamma \Gamma_{p'} \in \mathcal{L}.
\end{equation}

The inducing unitary representation $D(\Lambda)$ can be represented as a direct integral of IURs, as explained in ref.\ \cite{Toller}. The matrix elements of these IURs, indicated by $D^{M c}_{jmj'm'}(\Lambda)$, are described in refs.\  \cite{GGV,Naimark,Ruhl}. The possible values of the indices are
\begin{eqnarray}
&M = 0, \pm 1, \pm 2,\ldots, \qquad  c^2 < 1,&
\nonumber \\
&j = |M|, |M| + 1,\ldots, \qquad  m = -j, -j + 1,\ldots, j. \qquad &
\end{eqnarray}
If $M \neq 0$, $c$ must be imaginary. The representations $D^{Mc}$ and $D^{-M, -c}$ are unitarily equivalent. Moreover, there is the trivial one-dimensional representation, which we indicate by $D^{01}$. The restriction of these representations to the rotation subgroup is given by
\begin{equation} 
D^{M c}_{jmj'm'}(\Theta) = \delta_{jj'} R^j_{mm'}(\Theta), \qquad \Theta \in SO(3).
\end{equation}

In order to simplify the formalism, avoiding integrals over the parameter $c$, we assume that $D$ is a direct sum of IURs of the kind described above, labeled by the index $\gamma$. This means that the elements of $\tilde{\mathcal{H}}$ are described by the wave function $\Psi_{\gamma jm}(p)$ with the norm given by
\begin{equation} 
\|\Psi\|^2 = \int_{\mathcal{Q}} \sum_{\gamma jm} |\Psi_{\gamma jm}(p)|^2 \, d\nu(p).
\end{equation}
and the induced representation (\ref{Ind2}) takes the form
\begin{equation} \label{Ind3}
[S(\Gamma) \Psi]_{\gamma jm}(p) = D^{M c}_{jmj'm'}(\Lambda) \Psi_{\gamma j'm'}(p'),
\end{equation}
where the sum over the indices $j', m'$ is understood and the quantities $M, c$ depend on the index $\gamma$.

The GWF (\ref{Wig}) takes the more explicit form
\begin{eqnarray} 
&\rho(x) = (2 \pi)^{-4} \int \exp(i k \cdot x) \overline{\Psi_{\gamma jm}(p)}
D^{M c}_{jmj'm'}(\Lambda)& 
\nonumber \\ \label{Wig2}
&\times \Psi_{\gamma j'm'}(p') \, d\nu(p) \, d^4 k,&
\end{eqnarray} 
where $p'$ and $\Lambda$ are given by eq.\ (\ref{IndDef}) and
\begin{equation} 
\Gamma = \exp(-l k^{\alpha} \Xi_{\alpha 5}).
\end{equation} 

The generators $\tilde M_{\rho \sigma}$ defined by eq.\ (\ref{Inf}) can be decomposed into a part which acts on the angular momentum indices $j, m$ and a part which acts on the dependence of the wave function on $p$, in particular we can write
\begin{equation} \label{XTilde}
[\tilde X^{\alpha}\Psi]_{\gamma jm}(p) = Z^{\gamma\alpha}_{jm j'm'}(p) \Psi_{\gamma j'm'}(p) 
+ Y^{\alpha} \Psi_{\gamma jm}(p),
\end{equation}
where the matrix $Z^{\gamma\alpha}_{jm j'm'}(p)$ is Hermitian and
\begin{equation} \label{DiffOp}
Y^{\alpha} = -i l \left( \xi^5 \frac{\partial}{\partial \xi_{\alpha}} - 
\xi^{\alpha} \frac{\partial}{\partial \xi_5} \right).
\end{equation}
These derivatives act on an arbitrary smooth extension of a function in a neighborhood of $\mathcal{Q}$ in $\mathbf{R}^5$.

We consider the operators $f$ acting on the space $\tilde{\mathcal{H}}$ by multiplying the wave function by a function $f(p)$. We assume that $f(p)$ is bounded and infinitely differentiable, but it may be useful to assume that it has different properties. The operator $f$ satisfies the commutation relations
\begin{equation}
[\tilde M^{\mu \nu}, f] = i \left( \xi^{\mu} \frac{\partial}{\partial \xi_{\nu}} - 
\xi^{\nu} \frac{\partial}{\partial \xi_{\mu}} \right)f.
\end{equation}
In particular we have
\begin{equation} \label{Com}
[\tilde X^{\alpha}, f] = -i l \left( \xi^5 \frac{\partial}{\partial \xi_{\alpha}} - 
\xi^{\alpha} \frac{\partial}{\partial \xi_5} \right)f.
\end{equation}

\section{The intertwining operator.}

First we have to define the set $\mathcal{V} \subset \mathcal{Q}$, which contains the physical values of the energy-momentum $p$. It is a union of orbits of $\mathcal{Q}$ with respect to the action of $\mathcal{L}$, which are isomorphic, as homogeneous spaces, to the mass-shell of a massive particle. On these orbits we must have $\xi_{\alpha} \xi^{\alpha} > 0$ and  $\xi^5$ has a constant value $|\xi^5| > 1$. In order to have a connected set, we also require $\xi^0 > 0$ and $\xi^5 > 1$. 

On $\mathcal{V}$ one can use the coordinates $\xi^{\alpha}$ which have simple transformation properties under the group $\mathcal{G}$, but, in order to describe the space-time translations by means of eq.\ (\ref{Trans1}), one has to introduce the coordinates $p^{\alpha}$. Since both the coordinates transform as four-vectors under the Lorentz group, we must have
\begin{equation} \label{PXi}
((\xi^5)^2 -1)^{-1/2} \xi^{\alpha} = s^{-1/2} p^{\alpha}
\end{equation}
and the relation between the two coordinate systems is determined by the increasing function $\xi^5(s)$. In the absence of massless particles, we have $\xi^5(s) \geq \xi^5(s_0) > 1$. The measure $\nu$, restricted to $\mathcal{V}$, can be written in the form
\begin{eqnarray}
&\nu(p) = J(s) \, d^4 p,& 
\nonumber \\
&J(s) = 2 l^{-4} s^{-1} |(\xi^5)^2 - 1| \left|\frac{d\xi^5}{ds}\right|.&
\end{eqnarray}

Note that different choices of the function $\xi^5(s)$ define different observables $X^{\alpha}$ on the same system, described by a given Hilbert space $\mathcal{H}$ and a given unitary representation $U$ of the Poincar\'e group.

Each of the orbits we have chosen contains a rotation invariant point $\hat p(s)$ with coordinates 
\begin{equation} 
\boldsymbol{\xi} = 0, \qquad \xi^5 = \cosh\eta, \qquad \xi^0 = \sinh\eta, \qquad \eta > 0
\end{equation}
and we can write
\begin{equation} 
\hat p(s) = \exp(\eta \, \Xi_{50}) \hat p,
\end{equation}
where $\hat p$ is the point introduced in the preceding section. In agreement with eq.\ (\ref{Boosts}), we write the other points of the orbits in the form
\begin{equation} \label{Gamma}
p = \Lambda_p \hat p(s) = \Gamma_p \hat p, \qquad \Gamma_p = \Lambda_p \exp(\eta \, \Xi_{50}).
\end{equation}
In this way, we have partially determined, for $p \in \mathcal{V}$, the choice of the elements $\Gamma_p$ introduced in eq.\ (\ref{Boosts2}). 

The intertwining operator $A$ is not uniquely determined, because the same system can define different events. For instance, a many particle system defines several events corresponding to the collision of different pairs of particles. Important constraints are the Lorentz and the translation symmetries given by eqs.\ (\ref{LorCov}) and (\ref{AMu}). The second equation implies that $A$ is diagonal with respect to the variable $p$. Since the wave function $\psi(p)$ is not defined outside $\mathcal{V}$, we have
\begin{eqnarray}
&\Psi(p) = [A\psi](p) = 0, \qquad p \notin \mathcal{V},&
\nonumber \\
&\Psi(p) =  J^{-1/2}(s) A(p) \psi(p), \qquad p \in \mathcal{V}, \qquad &
\end{eqnarray}
where the bounded linear operator $A(p)$ is represented by a matrix which acts on the indices of the wave function $\psi(p)$ (not written explicitly).

From eqs.\ (\ref{Ind}), (\ref{LorCov}) and (\ref{Ind2}), using the convention (\ref{Gamma}) for $\Gamma_p$, we obtain
\begin{eqnarray} 
&A(p) R_s(\Lambda_p^{-1} \Lambda \Lambda_{p'}) = 
D(\Lambda_p^{-1} \Lambda \Lambda_{p'}) A(p'),&
\nonumber \\
&\Lambda \in \mathcal{L}, \qquad p' = \Lambda^{-1} p.&
\end{eqnarray}
For $\Lambda = \Lambda_p$, we have  $p' = \hat p(s)$, $\Lambda_{p'} = 1$ and
\begin{equation} 
A(p)  =  A(\hat p(s)) = A(s). 
\end{equation}
Then we obtain the condition
\begin{equation} 
A(s) R_s(\Theta) = D(\Theta) A(s), \qquad \Theta \in SO(3),
\end{equation}
which implies that the matrix which represents the operator $A(s)$ is diagonal with respect to the indices $j, m$ and does not depend on $m$. In conclusion, we have
\begin{equation} 
\Psi_{\gamma jm}(p)  = J^{-1/2}(s) A_{\gamma\sigma}^j(s) \psi_{\sigma jm}(p). 
\end{equation}
Eq.\ (\ref{Isom}) gives the condition
\begin{equation} 
\overline{A_{\gamma\sigma'}^j(s)} A_{\gamma\sigma}^j(s) = \delta_{\sigma\sigma'}
\end{equation}
(no sum over the index $j$).

Under certain conditions, namely when the wave function $\psi$ vanishes outside a region where $|\xi^\alpha| \ll 1$, we must recover the results of ref.\ \cite{Toller}.  In fact, if we put, in the relevant region, 
\begin{equation} \label{Simple}
\xi^{\alpha} = l p^{\alpha}, \qquad
\xi^5 = (l^2 s + 1)^{1/2},
\end{equation}
we have
\begin{eqnarray} 
&J(s) \approx 1, \qquad \eta \approx l s^{1/2},& 
\nonumber \\
&p^{\prime \alpha} \approx  p^{\alpha} - k^{\alpha}, \qquad
\Lambda \approx \Lambda_p^{-1} \Lambda_{p'}&
\end{eqnarray}
and eq.\ (\ref{Wig2}) takes the form derived in ref.\ \cite{Toller}.

In this approximation, the GWF is positive and it defines a POVM on the space-time. If in eq.\ (\ref{Wig2}) we substitute
\begin{equation} 
\psi(p) \to [T(a) \psi](p) = \exp(i a \cdot p) \psi(p),
\end{equation} 
we obtain
\begin{equation} 
\rho(x) \to \rho(x - a),
\end{equation} 
namely the GWF and the corresponding POVM are covariant under the translation group in agreement with eq.\ (\ref{Cov}). 

\section{Translations.}

It is known that the Snyder model has some problems with covariance under space-time translations. In the present paper we assume the existence of the unitary representation $U(a, \Lambda)$ of the Poincar\'e group acting on the physical Hilbert space $\mathcal{H}$. According to Wigner's treatment of symmetry operations \cite{Wigner,Wigner3,Bargmann}, this follows from the general rules of quantum mechanics and from the existence of equivalent classical reference frames. If we realize that only quantum reference frames exist \cite{AK,Rovelli,Toller2,Mazzucchi}, the situation may be different, but a complete consistent treatment of quantum frames is not yet available.  We have already discussed the Lorentz transformations properties of the coordinate operators $X^{\alpha}$ and of the corresponding GWF. Now we deal with the translations. 

We indicate by $f$ a function of the variables $\xi^{\alpha}$ defined on $\mathcal{V}$ and the correpsonding multiplication operator acting on the space $\mathcal{H}$. From eq.\ (\ref{Com}) and the properties of the operator $A$ we obtain
\begin{equation} \label{CommuXf}
[X^{\alpha}, f] = -i l \xi^5 \frac{\partial f}{\partial \xi_{\alpha}}.
\end{equation}
In particular we have
\begin{equation} 
[X^{\alpha}, p^{\beta}] = -i l \xi^5 \frac{\partial p^{\beta}}{\partial \xi_{\alpha}}
\end{equation}
and the following relation between  the coordinates measured in a given frame and in another translated frame
\begin{eqnarray} \label{TransX}
&T(-a) X^{\alpha} T(a) = \exp(-i a \cdot p) X^{\alpha} \exp(i a \cdot p)&
\nonumber \\
&= X^{\alpha} + l a^{\beta} \xi^5 \frac{\partial p_{\beta}}{\partial \xi_{\alpha}}.&
\end{eqnarray}
If we adopt eq.\ (\ref{Simple}), we obtain the simpler formula
\begin{equation} 
T(-a) X^{\alpha} T(a) = X^{\alpha}  + (l^2 s + 1)^{1/2} a^{\alpha}.
\end{equation}

We see that the new coordinates depend on the old coordinates and on the energy-momentum of the object that defines the event (more precisely, one should speak of the quantum averages of these quantities). If the averages of the coordinates of two events, defined by two objects with different energy-momenta, coincide when observed by a given frame, in general they do not coincide when observed by a translated frame. In other words, the space-time coincidence of events is not an absolute concept, in the same way as time coincidence (simultaneity) is not an absolute concept in special relativity. Einstein \cite{Einstein} stressed that the absolute character of space-time coincidence is one of the fundamental principles of general relativity. However, it should not be considered as a dogma. 

It is easy to see that, for any choice of the function $\xi^5(s)$, equation (\ref{TransX}) is experimentally wrong when applied to a macroscopic object for which $l^2 s$ is not negligible compared to 1. A related ambiguity appears if we consider a system composed of two noninteracting subsystems and only the first subsystem is used to determine the coordinates of the event. Then it is not clear if the quantities which appear on the right hand side of eq.\  (\ref{TransX}) concern the first subsystem or the whole system. Clarifying this ambiguity is preliminary for a treatment of macroscopic objects.

In order to avoid these problems, while waiting for some improvement of the formalism, one can restrict one's attention to events defined by few-particle systems. These difficulties are also present in other theories, like the ``doubly special relativity'' and related models \cite{AC}.  

It is useful to remark that, if we consider a single coordinate or a linear combination of the kind  $k \cdot X = k_{\alpha} X^{\alpha}$, we can always find an operator $F$, depending on $k$, defined by a function $F(p)$ with the properties
\begin{equation} 
[F, k \cdot X] = i,
\end{equation} 
\begin{equation} \label{TrasKX}
\exp(-i\lambda F) k \cdot X \exp(i\lambda F) = k \cdot X + \lambda.
\end{equation} 
Since the formalism is Lorentz symmetric, we may consider in detail only the coordinates $X^0$ and $X^1$.

In the first case we choose on $\mathcal{V}$ the coordinates $\boldsymbol{\xi}$ and $\alpha$ defined by
\begin{equation} \label{Alpha}
\alpha = \ln(\xi^0 + \xi^5) > \ln(1 + \|\boldsymbol{\xi}\|) = \hat\alpha
\end{equation}
and we have
\begin{equation} \label{FZero}
[F_0, X^0] = i, \qquad F_0 = l^{-1} \alpha.
\end{equation} 

In the other case, we choose on $\mathcal{V}$ the coordinates  $\xi^0$, $\xi^2$, $\xi^3$ and $\beta$ defined by
\begin{eqnarray} 
&\xi^1 = v \sin\beta, \qquad \xi^5 = v \cos\beta, \qquad
v = (1 + w^2)^{1/2},&
\nonumber \\ \label{Beta}
&|\beta| < \arccos(v^{-1}) = \arctan w =\hat\beta, &
\end{eqnarray} 
where 
\begin{equation} 
w^2 = (\xi^0)^2 - (\xi^2)^2 - (\xi^3)^2
\end{equation} 
and we obtain 
\begin{equation} 
[F_1, X^1] = i, \qquad F_1 = -l^{-1} \beta.
\end{equation} 

A similar treatment can be given if we choose the expression (\ref{Coord2}) for the space coordinates. In this case, however, we also find
\begin{equation} 
[F_0, X^r] = 0, \qquad r = 1, 2, 3,
\end{equation} 
where $F_0$ is given by eq.\ (\ref{FZero}), namely one can define a time translation which acts in the usual way on all the four coordinates. However, one cannot define space translations with the same property. The better behavior under time translations is paid by a worse behavior under Lorentz boosts.

Note that eq.\ (\ref{TrasKX}) gives a transformation property of the non-self-adjoint operator $k \cdot X$ and of the average value of the corresponding observable. A complete description of the statistical properties of this observable, however, requires knowledge of the corresponding POVM defined by eq.\  (\ref{POVM}). We may expect a covariance property of the kind
\begin{equation} \label{CovKX}
\exp(-i\lambda F) \mu_k(I) \exp(i\lambda F) = \mu_k(I - \lambda),
\end{equation} 
but this equation does not follow from eq.\ (\ref{TrasKX}). In fact, if the four-vector $k$ is spacelike, the self-adjoint operator  $k \cdot \tilde X$ generates a rotation and has a discrete spectrum. This means that the support of the spectral measure $\tilde \mu_k$, and also of the POVM $\mu_k$, is a discrete subset of the real line, in contradiction with eq.\  (\ref{CovKX}). This remark shows how delicate is the treatment of observables not described by self-adjoint operators.

\section{Variance of the coordinate observables.}

Now we derive some lower bounds for the variances given by eq. (\ref{Var}). These bounds must hold for any choice of the physical wave function $\psi \in \mathcal{H}$ and of the intertwining operator $A$ and we may more simply require that they hold for any choice of the wave function $\Psi \in \tilde{\mathcal{H}}$, provided that it vanishes outside the region $\mathcal{V}$. We also assume $s_0 = (2 m_0)^2 = 0$. It follows that our results do not depend on the choice of the intertwinig operator $A$ and on the relation between $\xi^5$ and $s$ introduced in section V.

We consider first a particular class of events defined by the head-on collision of two spinless particles, even if it is physically rather difficult to prepare an high-energy state of this kind. Then we extend the results to arbitrary events. In this simple case, the center-of-mass angular momentum is $j = 0$ and the index $\sigma$, which represents the center-of-mass helicities, takes only one value. We also assume that $D$ in eq.\ (\ref{Ind2}) is the trivial one-dimensional representation. It is shown in ref.\ \cite{Toller} that, in a commutative space-time, this means that the event is \textit{quasi-baricentric}, namely it defines a point which is as near as possible to the world line of the center-of-mass of the object, compatibly with the quantum uncertainty relations. 

The wave functions have no indices and we can write eq.\ (\ref{Wig2}) in the simpler form
\begin{equation} \label{Wig3}
\rho(x) = (2 \pi)^{-4} \int \exp(i k \cdot x) \overline{\Psi(p)} 
\Psi(p') \, d\nu(p) \, d^4 k.
\end{equation} 
From eq.\ (\ref{XTilde}) we see that $\tilde X^{\alpha} = Y^{\alpha}$, where $Y^{\alpha}$ is given by eq.\ (\ref{DiffOp}) and from the results of section III we obtain
\begin{eqnarray} 
&(\Delta x^{\alpha})^2 =
\langle (x^{\alpha} - c^{\alpha})^2 \rangle  = 
\int |[(Y^{\alpha} - c^{\alpha})\Psi](p)|^2  \, d\nu(p),&
\nonumber \\
&c^{\alpha} = \langle x^{\alpha} \rangle.&
\end{eqnarray}

First we consider $\Delta x^0$. Introducing the variables described in eq.\ (\ref{Alpha}), we have
\begin{equation}
Y^0 = -il \frac{\partial}{\partial \alpha}, \qquad d \nu(p) = d^3 \boldsymbol{\xi} \, d \alpha.
\end{equation} 
We use the following family of wave functions parametrized by the variable  $\lambda > 0$ 
\begin{equation}
\Psi_{\lambda}(\boldsymbol{\xi}, \alpha) = 
C_{\lambda} \exp(i l^{-1} c \alpha) f(\boldsymbol{\xi}, \lambda (\alpha - \hat \alpha)),
\end{equation}  
where $f$ is a smooth function with compact support vanishing for negative values of its second argument. We have, after the change of variable $t = \lambda\alpha$,
\begin{equation}
1 = \int |\Psi(p)|^2 \, d\nu(p) = 
|C_{\lambda}|^2 \lambda^{-1}\int |f(\boldsymbol{\xi}, t - \lambda \hat\alpha)|^2 \, 
d^3 \boldsymbol{\xi} dt,
\end{equation}  
\begin{equation}
\langle (x^0 - c)^2\rangle = |C_{\lambda}|^2 l^2 \lambda \int |f'(\boldsymbol{\xi}, t - \lambda \hat\alpha)|^2 \, d^3 \boldsymbol{\xi} dt,
\end{equation} 
where $f'$ is the derivative of $f$ with respect to its second argument. It follows immediately that
\begin{equation}
\lim_{\lambda \to 0} \langle (x^0 - c)^2\rangle = 0,
\end{equation} 
namely that $\Delta x^0$ can take arbitrarily small values.

Then we consider $\Delta x^1$ and use the variables described in eq.\ (\ref{Beta}). We have
\begin{equation} 
Y^1 = i l \frac{\partial}{\partial\beta}, \qquad
d \nu(p) = d \xi^0 \,d \xi^2 \,d \xi^3 \, d \beta.
\end{equation}

If $f(\beta)$ is a continuos piecewise differentiable function which vanishes for $|\beta| \geq \hat\beta$, one can prove, by means of the standard methods of variational calculus, the inequality 
\begin{equation} \label{Ineq}
\int |f'(\beta)|^2 \, d \beta \geq \pi^2 (2 \hat\beta)^{-2}\int |f(\beta)|^2 \, d \beta.
\end{equation}
The equality holds for
\begin{equation} 
f(\beta) = C \cos\frac{\pi\beta}{2 \hat\beta}.
\end{equation}

If we put
\begin{equation}
\Psi(p) = \exp(-i l^{-1} c \beta) f(\xi^0, \xi^2, \xi^3, \beta),
\end{equation}  
from this inequality we obtain
\begin{eqnarray} 
&\langle (x^1 - c)^2\rangle = l^2 \int |f'(\xi^0, \xi^2, \xi^3, \beta)|^2 \, d \nu(p)&
\nonumber \\
&\geq \pi^2 l^2 \int |\Psi(p)|^2 (2 \hat\beta)^{-2} \, d \nu(p).& 
\end{eqnarray}
The quantity $\hat\beta$ approaches its upper bound $\pi/2$ when $w$ is very large and we obtain
\begin{equation} 
\langle (x^1 - c)^2\rangle > l^2. 
\end{equation}
One approaches this lower bound if the wave function has the form
\begin{equation} \label{Minim}
\Psi(p) = \exp(-i l^{-1} c \beta) f(\xi^0, \xi^2, \xi^3) \cos \beta
\end{equation}
and $f$ is negligible unless $w \gg 1$.
Similar inequalities hold for the other space coordinates and we obtain
\begin{equation} 
\Delta x^r > l, \qquad r = 1, 2, 3.
\end{equation}

Now we have to show that one cannot obtain smaller variances by using objects of a more general kind. We start from the formula
\begin{eqnarray} 
&\langle (x^1 - c)^2 \rangle = 
\int \sum_{\gamma jm} |Z^{\gamma 1}_{jm j'm'}(p) \Psi_{\gamma j'm'}(p)&  
\nonumber \\
&+ (Y^1 - c) \Psi_{\gamma jm}(p)|^2 \, d\nu(p). &
\end{eqnarray}
We put
\begin{equation} 
\Psi_{\gamma jm}(p)  = \exp(-i l^{-1} c \beta) 
U^{\gamma}_{jm j'm'}(p) \tilde \Psi_{\gamma j'm'}(p),
\end{equation}
where $U^{\gamma}_{jm j'm'}(p)$ is a unitary matrix with the property
\begin{equation} 
\frac{\partial}{\partial\beta}  U^{\gamma}_{jm j'm'}(p) = i l^{-1} Z^{\gamma 1}_{jm j''m''}(p)
U^{\gamma}_{j''m'' j'm'}(p).
\end{equation}
We obtain
\begin{equation} 
\langle (x^1 - c)^2 \rangle  =
\int \sum_{\gamma jm} |Y^1 \tilde \Psi_{\gamma jm}(p)|^2 \, d\nu(p)
\end{equation}
and we can use the inequality (\ref{Ineq}) as in the simple case to get the required result
\begin{equation} 
\langle (x^1 - c)^2 \rangle  \geq \pi^2 l^2
\int \sum_{\gamma jm} |\tilde \Psi_{\gamma jm}(p)|^2 (2 \hat\beta)^{-2} \, d\nu(p) > l^2.
\end{equation} 

If we take into account the Lorentz symmetry of the formalism, we can write the result in the form
\begin{equation} 
\Delta (k \cdot x) > l |k \cdot k|^{1/2} \theta(- k \cdot k),
\end{equation}
where $\theta$ is the step function. Note that the right hand side is continuous when the four-vector $k$ crosses the light cone. This formula describes completely the possible values of the variance of a single coordinate.

One may ask how an observable which has a discrete spectrum, for instance $k \cdot X$ with k spacelike, has a lower bound for the dispersion. This happens because the probability  $(\psi, \tau_k (\{\lambda\})\psi)$, where $\lambda$ is a point of the spectrum, cannot approach the value $1$. In fact, the vectors in the range of the projection operator $\tilde \tau(\{\lambda\})$ have an unphysical energy-momentum spectrum.  In other words, we have $\|\tau(\{\lambda\})\| < 1$ and the POVM $\tau$ corresponding to the observable $k \cdot X$ does not possess the ``norm-1-property'' discussed in ref.\ \cite{HLPPY}.
 
There are also inequalities that involve the dispersions of two or more coordinates, measured separately on systems in the same state, namely prepared in the same way. We give only two simple examples concerning head-on collisions and we assume that $c^{\alpha} = \langle x^{\alpha} \rangle = 0$. We consider first a wave function of the kind (\ref{Minim}), which permits $\langle (x^1)^2 \rangle$ to approach its lower bound $l^2$, and we compute the quantity
\begin{eqnarray} 
&\langle (x^0)^2 \rangle  =
\int |Y^0 f(\xi^0, \xi^2, \xi^3) \cos \beta|^2 \, d\nu(p)&
\nonumber \\
&= l^2 \int \left|\frac{\partial f}{\partial \xi^0} v (\cos \beta)^2 + 
f v^{-1} \xi^0 (\sin \beta)^2 \right|^2
 \, d \xi^0 \,d \xi^2 \,d \xi^3 \, d \beta&
\nonumber \\
&= l^2 2^{-3} \pi \int \left(3 \left|\frac{\partial f}{\partial \xi^0}\right|^2 v^2 + 
|f|^2 (3 v^{-2} (\xi^0)^2 -1) \right) \, &
\nonumber \\
&\times d \xi^0 \,d \xi^2 \,d \xi^3 > 2^{-1} l^2. &
\end{eqnarray}
We have performed a partial integration with respect to $\xi^0$ and the integration with respect to $\beta$. We see that $\langle (x^0)^2 \rangle$ and $\langle (x^1)^2 \rangle$ cannot both approach their lower bounds in the same state.

We also consider the wave function
\begin{equation} 
\Psi(p) = f(\xi^0) \xi^5, 
\end{equation} 
where $f$ is negligible unless $\xi^0 \gg 1$. With some calculations, we obtain
\begin{equation} 
\langle (x^r)^2\rangle \approx l^2, \qquad r = 1, 2, 3, \qquad
\langle (x^0)^2\rangle \approx 2 l^2.
\end{equation}
We see that all three quantities $\langle (x^r)^2\rangle$ can approach their lower bounds in the same state, but in this case $\langle (x^0)^2\rangle$ cannot be too small.

\section{Joint measurement of the coordinates.}

In the preceding sections we always treat measurements of a single coordinate or of a linear combination of them $k \cdot X$.  When we consider the uncertainty relations involving two coordinates, it is understood that they refer to measurements performed on two different systems, prepared in the same way. However, for a complete physical interpretation, it is necessary to consider approximate joint measurements of different coordinates on the same system. 

The same problem appears when we consider the joint measurement of the canonical coordinates $p$ and $q$ in the phase space. This problem was treated in ref.\ \cite{LT} by using essentially the Wigner function \cite{Wigner2}, which unfortunately was not called by name there. A related treatment, with a deeper discussion of the physical motivations, is given in ref.\ \cite{Busch}. 

The approach of  ref.\ \cite{LT} starts from a formulation of the correspondence principle in terms of observables with two possible outputs, called \textit{effects} \cite{Ludwig} or \textit{tests} \cite{Giles}. They can be considered as POVMs defined on a set composed of only two points, say $\{1, 0\}$, the corresponding positive operators being $F$ and $1-F$. In a classical (non-quantum) theory a test is described by a continuous function $0 \leq f(p, q)\leq 1$ defined on the phase space, which gives the probability of obtaining the result $1$ if the state is represented by the point $(p, q)$ of the phase space. We are considering for simplicity a system with one degree of freedom.

The problem is to find the positive operator $F$ which corresponds to the positive function $f$ and a natural solution is provided by the Weyl rule \cite{Weyl}
\begin{equation} 
F = (2\pi)^{-1} \int \tilde f(\sigma, \tau) \exp(-i\sigma P + i\tau Q) \, d \sigma \, d \tau,
\end{equation}
\begin{equation} 
\tilde f(\sigma, \tau) = (2\pi)^{-1} \int f(p, q) \exp(i\sigma p - i\tau q) \, d p \, d q,
\end{equation}
where $P$ and $Q$ are the quantum operators corresponding to the canonical coordinates $p$ and $q$. The operator $F$ obtained in this way is not necessarily positive and, in fact, the classical tests which determine a point of the phase space with too high a precision have no corresponding quantum test.

The quantum probability of obtaining the result $1$ if the state is defined by the vector $\psi$ is given by
\begin{equation} 
(\psi, F \psi) = \int  f(p, q) \rho(p, q) \, d p \, d q,
\end{equation}
where
\begin{eqnarray} 
&\rho(p, q) = (2\pi)^{-2}\int \exp(i\sigma p - i\tau q) &
\nonumber \\
&\times (\psi, \exp(-i\sigma P + i\tau Q) \psi) \, d \sigma \, d \tau &
\end{eqnarray}
is the Wigner function. 

The analogy with eq.\ (\ref{Wig}) is evident: the representation $S$ of $\mathcal{G}$ is replaced by the representation $\exp(-i\sigma P + i\tau Q)$ of the Weyl-Heisenberg group, which can also be considered as a projective (ray) representation of the translation group of the phase space \cite{Weyl}. The analogy can be carried further. A classical test which describes an approximate measurements of the four space-time coordinates is represented by a continuous function $0 \leq f(x)\leq 1$ defined in the classical Minkowski space-time. The corresponding quantum test, if it exists, is described by a positive operator $F$ defined by
\begin{equation} \label{QTest}
(\psi, F \psi) = \int  f(x) \rho(x) \, d^4 x,
\end{equation}
where the GWF $\rho$ is given by eq.\ (\ref{Wig}). It is positive only if the function $f$ has suitable properties.

If we consider a test describing an approximate measurement of the single coordinate $k \cdot x$, in the classical theory it is described by the positive function $f(k \cdot x)$ and in the quantum theory by the positive operator
\begin{equation} 
F = \int  f(\lambda) \, d \tau_k(\lambda),
\end{equation}
in agreement with the statistical meaning of the POVM $\tau_k$ introduced in section IV. In this case, eq.\ (\ref{QTest}) follows from the treatment of section IV and the operator $F$ is automatically positive. The complete characterization of the functions $f(x)$ which correspond to positive operators $F$ is a difficult problem.

\section{Conclusions.}

We have examined Snyder's model of non-commutative space-time from a particular point of view, namely by considering the space-time coordinates as ordinary quantum observables describing measurements on the physical object which determines an event.  The final aim is an approximate description of some of the effects of quantum gravity, when the average value of the gravitational field is negligible. 

There are several alternative models and we do not claim to have chosen the best one.  However, we think that some of the following ideas, developed in the preceding sections, can be applied to a large class of models. 

\begin{itemize}

\item The model does not directly require any deformation of the Poincar\'e group and of its Lie algebra, although it is not excluded. The only deformed objects are the operators $X^{\alpha}$ which describe the space-time coordinates.

\item As in the commutative theory, the spectral condition requires that the operators $X^{\alpha}$ cannot be self-adjoint and their statistical properties must be described by means of a POVM acting on the physical Hilbert space $\mathcal{H}$. It is obtained, by means of an intertwining operator, from a spectral measure acting on an auxiliary Hilbert space $\tilde{\mathcal{H}}$. In the class of models we are considering, it is the spectral measure of a generator of a representation $S$ of a group $\mathcal{G}$ containing the Lorentz group.

\item The usual covariance property of the coordinates with respect to space-time translations has to be modified. We propose to interpret this feature as a break-down of the absolute character of the concept of space-time coincidence, which is one of the foundations of classical general relativity.

\item From the algebraic properties of the model and the properties of the energy-momentum spectrum, one can derive lower bounds to the variance of the coordinate observables. These inequalities are strongly model dependent.

\item The model confirms that there is no contradiction between Lorentz symmetry and limitations to the accuracy of length measurements or a discrete spectrum of the coordinate observables \cite{RS}.

\item The model provides an example of a POVM that does not possess the ``norm-1-property'' discussed in ref.\ \cite{HLPPY}.

\item Several different definitions of the coordinate observables may coexist in the same quantum theory, if we do not introduce any limitation to the positive bounded operators that can describe physical observables, in particular the tests $\tau_k(I)$. In a satisfactory formalism, the choice of the physically correct model of non-commutative space-time should follow from an accurate definition of the observables of the theory.

\end{itemize}

\end{document}